%% file: main.tex
\documentclass[conference]{IEEEtran}
\IEEEoverridecommandlockouts
\usepackage{amsmath,amssymb,amsfonts}
\usepackage{algorithmic}
\usepackage{graphicx}
\usepackage{textcomp}
\usepackage{amsmath,amssymb,amsfonts}
\usepackage{algorithmic}
\usepackage{graphicx}
\usepackage{textcomp}
\usepackage{xcolor}
\usepackage{tikz}
\usepackage[edges]{forest}
\usepackage{array}
\newcolumntype{C}[1]{>{\centering\arraybackslash}p{#1}}
\newcolumntype{L}[1]{>{\raggedright\arraybackslash}p{#1}}
\usetikzlibrary{arrows.meta,shadows}
\usepackage{enumitem}   
\usepackage{tabularx}
\usepackage{booktabs}   
\usepackage{caption}    
\usepackage{graphicx}
\usetikzlibrary{shapes.geometric, arrows, positioning, chains}
\usepackage{xcolor}
\usepackage{pgfplots}
\usepackage{enumitem}
\def\BibTeX{{\rm B\kern-.05em{\sc i\kern-.025em b}\kern-.08em
    T\kern-.1667em\lower.7ex\hbox{E}\kern-.125emX}}
\usepackage{cite}
\usepackage{multirow}

\author{
    \IEEEauthorblockN{
        Uchswas Paul\IEEEauthorrefmark{1},
        Ananya Mantravadi\IEEEauthorrefmark{1},
        Jash Shah\IEEEauthorrefmark{1},
        Shail Shah\IEEEauthorrefmark{1},\\
        Sri Vaishnavi Mylavarapu\IEEEauthorrefmark{1},
        M Parvez Rashid\IEEEauthorrefmark{2},
        Edward Gehringer\IEEEauthorrefmark{1}
    }
    \IEEEauthorblockA{\IEEEauthorrefmark{1}Department of Computer Science, North Carolina State University, Raleigh, NC, USA\\
    Emails: \{upaul, amantra, jshah23, sshah38, smylava, efg\}@ncsu.edu}
    \IEEEauthorblockA{\IEEEauthorrefmark{2}Department of Computer Science, College of Charleston, Charleston, SC, USA\\
    Email: rashidp@cofc.edu}
}

\begin{document}

\title{Scaling Success: A Systematic Review of Peer Grading Strategies for Accuracy, Efficiency, and Learning in Contemporary Education
}

\maketitle

\begin{abstract}

Peer grading has emerged as a scalable solution for assessment in large and online classrooms, offering both logistical efficiency and pedagogical value. However, designing effective peer-grading systems remains challenging due to persistent concerns around accuracy, fairness, reliability, and student engagement. This paper presents a systematic review of 122 peer-reviewed studies on peer grading spanning over four decades. Drawing from this literature, we propose a comprehensive taxonomy that organizes peer grading systems along two key dimensions: (1) evaluation approaches and (2) reviewer weighting strategies. We analyze how different design choices impact grading accuracy, fairness, student workload, and learning outcomes. Our findings highlight the strengths and limitations of each method. Notably, we found that formative feedback---often regarded as the most valuable aspect of peer assessment---is seldom incorporated as a quality-based weighting factor in summative grade synthesis techniques. Furthermore, no single reviewer weighting strategy proves universally optimal; each has its trade-offs. Hybrid strategies that combine multiple techniques could show the greatest promise. Our taxonomy offers a practical framework for educators and researchers aiming to design peer grading systems that are accurate, equitable, and pedagogically meaningful.

\end{abstract}

\vspace{3pt}

\section{Introduction}
\IEEEPARstart{P}{eer} grading has emerged as a promising alternative to traditional instructor-led evaluation methods, which, while ideal, are becoming increasingly difficult to sustain in large or open-format courses \cite{Wright2015}. This approach has gained traction, particularly with the rise of Massive Open Online Courses (MOOCs) and the growing class sizes \cite{Formanek2017}. Beyond reducing the grading burden on instructors~\cite{Price2016}, it also gives students an understanding of the requirements and alternative approaches to a problem \cite{Chiou2015, Wimshurst2013}.

Peer grading, though, is not without limitations. A recurring concern in the literature is the subjectivity and inconsistency of peer evaluations \cite{rashid2024navigating}, which may compromise the reliability and validity of the grades assigned by the peer \cite{Kravchenko2018}. Moreover, it can increase cognitive and temporal workload for the students \cite{Sridharan2019, Boudria2018}. Various approaches have been explored to mitigate these limitations. However, the different strategies have rarely, if ever, been compared, and the question remains: How can practitioners best minimize staff effort for reliable grading while maximizing learning gains for students?

Due to the increasing interest of the academic community in peer grading, there have been some systematic reviews of peer grading and assessment. However, many of these studies had some limitations. First, they had very limited scope. Panadero et al.\@\cite{Panadero17112019} conducted an empirical review on the effects of anonymity in peer assessment that included only 14 papers.  Misiejuk et al.\@\cite{MISIEJUK2021104319} conducted a scoping review of 10 papers that investigated different types of backward evaluation and the platforms used to facilitate that process. Other studies tend to focus on different aspects of peer grading without offering a comprehensive evaluation. Ravikiran \cite{ravikiran2020}, for instance, reviewed automatic grading tools and methods for detecting rough reviews. Gamage et al.\@\cite{Gamage03042021} offered a broad classification of peer assessment approaches, including submission strategies, reviewer types, and grading algorithms. However, their work was primarily quantitative, lacked qualitative synthesis, and focused solely on MOOCs, neglecting other educational contexts.

Overall, there remains a gap in the literature for a holistic, comparative analysis of peer grading methods---that evaluates their impact considering multiple factors such as accuracy, scalability, cognitive workload, and the learning outcomes for graders and authors.
To address this gap, our research centers on the following key questions:

\begin{itemize}
    \item \textbf{RQ1}: How do different peer-grading methods influence accuracy, reliability, and fairness?
    \item \textbf{RQ2}: How do various peer-grading approaches affect students' learning outcomes and graders' cognitive workload?
    \item \textbf{RQ3}: How can students and staff derive accurate peer grades with minimum effort?
\end{itemize}

\section{Materials and Methods}
Our systematic review followed the PRISMA framework \cite{PAGE2021790} to ensure rigor and clarity. The process began with clearly defined selection criteria focused on peer grading. A comprehensive search was conducted across scholarly databases using targeted keywords. After removing duplicates, studies were screened by title and abstract, followed by a full-text review. The final selection was analyzed to summarize current peer grading methods, identify gaps, and suggest directions for future research.

\subsection{Source of Information}
\label{source_of_information}
The data for this review were gathered from a range of reputable academic databases selected for their comprehensive coverage of scholarly literature across multiple disciplines. These sources include ACM Digital Library, Web of Science, EBSCO, ERIC, PsycINFO, IEEE Xplore, and PubMed. Each database was chosen for its focus on high-quality, peer-reviewed research and its relevance to the subject areas addressed in this review. By utilizing these well-established databases, the review ensured a broad and reliable selection of relevant studies, providing a solid foundation for the analysis and findings presented.

\begin{table}[h!]
\centering
\renewcommand{\arraystretch}{1.2}
\begin{tabular}{C{0.60\linewidth} C{0.30\linewidth}}  
\toprule
\textbf{Inclusion Criteria} & \textbf{Exclusion Criteria} \\
\midrule
\begin{itemize}[left=0pt, itemsep=0pt, parsep=0pt]
    \item Publications indexed in the following databases: Web of Science, ACM Digital Library, IEEE Xplore, PubMed, ERIC, PsycINFO, and EBSCO.
    \item Scientific articles, conference proceedings, PhD theses/dissertations, book chapters, and technical reports.
    \item Studies published between 1981 and 2024.
    \item Studies published in English language.
\end{itemize}
&
\begin{itemize}[left=0pt, itemsep=0pt, parsep=0pt]
    \item Duplicate studies.
    \item Books.
    \item Studies that do not address at least one research question.
    \item Studies with retraction notices or errata.
\end{itemize}
\\
\bottomrule
\end{tabular}
\captionsetup{font=small, labelfont=sc, justification=centering}
\caption{Inclusion and Exclusion Criteria for the \\ Systematic Literature Review.}
\label{tab:inclusion_exclusion}
\end{table}

\subsection{Eligibility Criteria}
The selection parameters for this review were defined to ensure relevance and quality. We included publications indexed in one of the seven databases mentioned in section \ref{source_of_information}, focusing on peer-reviewed journal articles, conference proceedings, book chapters, and technical articles.  To capture a comprehensive and diverse range of approaches within the peer-grading field, works published between 1981 and 2024 were included in the review. Duplicate studies, books, and retracted papers were discarded. The inclusion and exclusion criteria are listed in Table \ref{tab:inclusion_exclusion}.

\subsection{Search Strategy and Study Selection}

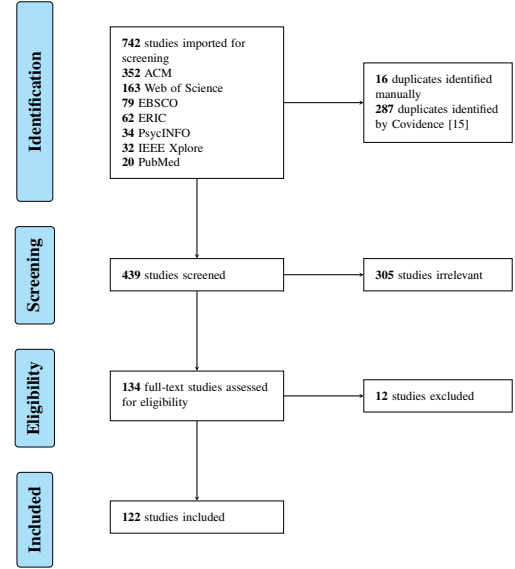
\begin{figure}[htbp]
    \centering
    \scalebox{0.33}{\input{prisma.tex}} 
    \caption{PRISMA flow diagram of the study selection process}
    \label{fig:prisma}
\end{figure}

An initial search was conducted in the selected databases using the keywords ``peer grading'' and/or ``peer marking'', combined with the predefined inclusion and exclusion criteria. This search returned a total of 742 publications. Of these, 303 were identified as duplicates, with 287 detected automatically by Covidence\cite{Covidence} and an additional 16 flagged manually by the authors. After removing duplicates, 439 papers remained and were screened by abstract to assess their relevance. Following this initial screening, 134 studies were selected for a full-text review. 12 papers were excluded because they did not address any of the research questions, resulting in a final selection of 122 studies for inclusion in the review (Figure \ref{fig:prisma}).

\subsection{Quality Control of Study}
Multiple authors were actively engaged throughout the research process to ensure the accuracy and trustworthiness of the studies included. Each study was independently screened and reviewed by two authors. Any disagreement was followed by collaborative discussions to achieve agreement and reduce the likelihood of individual or methodological bias influencing the process. In the screening phase, Cohen's Kappa ($\kappa$) was 0.54, suggesting ``moderate'' agreement \cite{cohens:kappa, Landis:Koch:Kappa:Range}, while in the eligibility checking phase, the authors achieved almost perfect agreement ($\kappa$=0.87).

\subsection{Taxonomy Development}
To systematically organize peer-grading systems, we developed a novel taxonomy based on insights from the literature. We followed a bottom-up approach to establish the taxonomy \cite{glaser1967discovery, nickerson2013method}. The first and second authors independently analyzed all peer-grading studies selected in the final review. Each rater carefully examined the studies' methodologies, strategy descriptions, and evaluation techniques. They independently identified features and strategies involved in peer-grading. The first author identified 11 distinct features of peer grading, while the second author identified 10 features, with eight features overlapping between them. These features were then refined, merged, or subdivided through iterative discussion, resulting in a consolidated set of features. Finally, the agreed-upon features were systematically organized into two orthogonal dimensions, providing a structured and comprehensive taxonomy of peer grading strategies. Any disagreements that arose were resolved by the last author, whose decisions were considered final.

\section{The Peer Grading System Taxonomy}

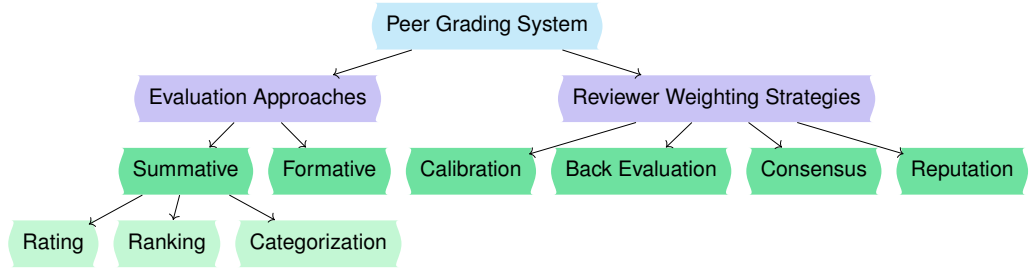
\begin{figure*}[ht]
    \centering
    \scalebox{0.8}{\input{peer_grading_diagram.tex}} 
    \caption{Taxonomy of Peer Grading Systems across Two Orthogonal Dimensions}
    \label{fig:peer_grading_framework}
\end{figure*}

Our proposed taxonomy, shown in Figure \ref{fig:peer_grading_framework}, organizes the peer-grading process into two orthogonal dimensions:  Evaluation Approaches and Reviewer Weighting Strategies. Together, those dimensions define how peer grading is conducted and processed. Below, we describe each dimension and its corresponding categories.

\subsection{Evaluation Approach}
The Evaluation Approach refers to how graders assess the student work. Graders can give summative and/or formative feedback in peer grading. We classify the evaluation approaches into two main categories: (1) Summative Feedback and (2) Formative Feedback.

\vspace{3pt}

 \subsubsection{Summative Feedback } Summative evaluation focuses on assigning a quantitative value based on the quality of work. Summative feedback can be given in three ways: Rating, Ranking, and Categorization.

\vspace{3pt}

\textit{\textbullet\ Rating}: Graders assign numeric scores to reflect their assessment of submission’s quality \cite{Arous2021, kay2022}. These scores can be raw points out of a total, as in the system proposed by Na et al.\@\cite{Na2019}, where graders assigned scores of up to 40 points for one question and 60 points for another. Often, grading is done using a predefined Likert scale---for example, a scale of 5, 10, or even 11 \cite{L.Zhang2023,Vu2018}. In many cases, students are provided with rubric-based criteria, and they must assign scores accordingly \cite{Vista2015, Bradley2019, Luaces2015}. Zhang et al.\@\cite{L.Zhang2023} defined a fixed rubric with categories such as Novelty, Function, Documentation, User-Friendliness, and Code Style, where graders gave scores on a 5 Likert scale on each item.

\vspace{3pt}
    
\textit{\textbullet\  Ranking}: Instead of providing a numeric score, graders rank a set of submissions relative to one another, creating an ordered list based on perceived quality \cite{Holston2018, Caragiannis2014, Moffat2020}. Graders are asked to make a partial ranking from a subset of submissions \cite{Raman2015, Han2018} or are sometimes asked to perform pairwise rankings by comparing two submissions \cite{ KRomić2023}. Some systems integrate ranking with rating. For example, Dery et al.\cite{L.Dery2024} developed a platform in which students first assign a rating to each submission and, whenever they give the same score to multiple submissions, are prompted to rank those tied works to refine their relative quality.

\vspace{3pt}

 \textit{\textbullet\ Categorization }: With categorization, graders assign submissions to predefined categories or levels based on specific criteria. One of the most common approaches is a binary response, where students select options from a checklist such as yes/no \cite{Joglekar2013, Kong2020, Storjohann2019}, positive/negative \cite{Zarkoob2019}, or high-quality/low-quality \cite{Dasgupta2013}. In some cases, more detailed scales are employed. For example, Chen et al.\@\cite{Chenwest2022} introduced a peer-grading system in which students assigned one of four predefined labels to each submission: high-level and correct, incorrect/incomplete, ambiguous/garbage, or too low-level.

\vspace{3pt}
\subsubsection{Formative Evaluation} Along with summative feedback, students often give formative feedback in the form of written comments. Formative feedback includes justification for the assigned score \cite{DBoud1981} and actionable recommendations to improve work \cite{Andriamiseza2023, deAlfaro2014, rashid2022going}.

\subsection{Reviewer Weighting Strategies}
\label{grade_weighting_technique}
Reviewer Weighting Strategies are methods used to refine or weight the influence of individual peer-assigned grades. Some systems adopt straightforward approaches such as unweighted average \cite{Kerr1995} or median \cite{Chakraborty2024} of the grades received. However, given the variability and potential noise in peer grading, many systems incorporate reviewer-specific reliability factors to weight individual contributions more effectively. Four commonly utilized methods for determining these weighting factors are: Calibration,  Back-evaluation, Consensus, and Reputation.


\vspace{3pt}

\subsubsection{Calibration-Based Reviewer Weighting}  Graders participate in a calibration exercise where they assess sample submissions with known scores. Their grading accuracy is evaluated, and their subsequent peer assessment scores are weighted according to their calibration performance \cite{Chakraborty2024, Chenwest2022, Fang2017, Wang2022}. The calibration phase can also serve as a training phase, helping reviewers understand and apply the assessment criteria effectively \cite{Badea2022}. 

\vspace{3pt}
    
\subsubsection{ Back-Evaluation Based Reviewer Weighting} Peer feedback is evaluated by its authors based on the quality of the feedback or grades they provide. These evaluations help determine the weight of each reviewer’s contribution to the final score. In some cases, direct teacher intervention is required if a grade is challenged by the assessee; the grade may be adjusted if necessary~\cite{Storjohann2019}. Sometimes, back-evaluation indirectly affects peer-grading weight. Reviewers gain reputation scores based on their reviews, which influence their impact on grade validation \cite{Alipour2022, Alipour2024}.  There are some back-evaluation scenarios where the reviewer’s grade for reviewing is either rewarded or penalized for providing good or poor reviews, respectively \cite{deAlfaro2014, EFGehringer2000}.

\vspace{3pt}
    
\subsubsection{ Consensus-Based Reviewer Weighting} The weight of a grader’s assessment is influenced by how closely it aligns with the evaluations of other reviewers. The more agreement there is with peers, the higher the weight and influence that grader’s score has on the final grade \cite{Hamer2005, Zarkoob2023, Sofjan2023}. In addition to rating- or numeric-based systems, consensus has also been applied to ranking-based evaluation systems. If a grader ranks submissions similarly to the emerging consensus, they are considered more reliable, and their rankings are weighted more in subsequent iterations \cite{Raman2015, Raman2014}. 

\vspace{3pt}

\subsubsection{ Reputation-Based Reviewer Weighting} This approach assigns different weights to grades based on the reviewer’s competency and historical performance, influencing those with a strong track record. A grader’s past grading behavior is used to assess their reputation. This can be determined in several ways: by considering past under-marking or over-marking tendency compared to peers \cite{Davies2006, Wang2019}, by evaluating how well a reviewer’s past grading aligns with instructors, or by considering both \cite{Zarkoob2024}. Some methods, like that of Walsh et al.\@\cite{Walsh2014}, used the grader’s own past assignment grades as a determinant. Subject-matter expertise can also serve as an indicator, for example, performance on chapter tests \cite{García-Martínez2019}. Additionally, participant engagement can influence reputation scores, such as the number of reviews completed \cite{Khatoon2022} or project deliverables submitted \cite{Badea2022}. 
    
\vspace{3pt}
Besides these four common factors, some studies also incorporate other factors to adjust the weight of assigned grades. For example, Namanloo et al.\@\cite{Namanloo2022} adjusted peer grades by accounting for social connections, reducing the influence of biased evaluations arising from friendships or conflicts of interest. Yuan et al.\@\cite{Yuan2020} assigned more weight to peer reviews that demonstrated higher thoroughness, as measured by the length and quality of the textual feedback.

Table~\ref{tab:reviewer-strategies} summarizes how many papers adopt each evaluation approach and reviewer weighting strategy.
\begin{table}[h!]
\centering
\begin{tabular}{llr}
\toprule
\textbf{Dimension} & \textbf{Category} & \textbf{Number of Papers} \\
\midrule
\multirow{4}{*}{Evaluation Approaches} 
  & Rating           & 95 \\
  & Ranking          & 17 \\
  & Categorization   & 10 \\
  & Formative & 62 \\
\midrule
\multirow{4}{*}{Reviewer Weighting Strategies} 
  & Calibration      & 21 \\
  & Back-Evaluation  & 11 \\
  & Consensus        & 27 \\
  & Reputation       & 22 \\
\bottomrule
\end{tabular}
\caption{Number of papers employing each evaluation approach and reviewer weighting strategy}
\label{tab:reviewer-strategies}
\end{table}

\section{Results}
This section presents a comparative analysis of each method in our taxonomy, structured around our research questions.

\subsection{RQ1: Impact on Accuracy, Reliability, Fairness}

While rating is predominantly used for evaluation, many systems have reported that ratings are often noisy, biased, and inconsistent \cite{Namanloo2022, Jimenez-Romero2013, Capuano2016}, requiring additional filtering or correction techniques to enhance their accuracy. Jimenez-Romero et al.\@\cite{Jimenez-Romero2013} evaluated the accuracy of peer-assigned scores compared to teacher-assigned scores, revealing an average error of 42\%. Formanek et al.\@\cite{Formanek2017} found rating to be highly unreliable and inconsistent, with an inter-rater agreement of just 0.32. Rating is also prone to grade inflation. Han et al.\@\cite{Han2023} observed that peer-assigned scores were often inflated in unregulated environments---sometimes by as much as 50 points compared to TA-assigned scores. These inaccuracies stem from several factors. For example, Steverding et al\@\cite{Steverding2016} showed that different student groups exhibit varying grading tendencies—some being more lenient, others harsher. Similarly, Luaces et al.\@\cite{Luaces2015} observed that some students consistently give generous grades while others are more critical. James et al.\@\cite{James2018} noted that graders often interpret rubrics differently, leading to inconsistent outcomes.  

The high variability and unreliability often reported in peer-assigned numeric scores can be mitigated by adopting a ranking-based evaluation system in place of rating. Several studies emphasize the greater reliability of rankings over absolute ratings~\cite{Ling2021, Raman2015}. Raman et al.\cite{Raman2014} argue that peer grading based on ordinal feedback is preferable, as it avoids requiring students to make cardinal judgments. Supporting this, Jingjing et al.\cite{Jingjing2015} demonstrated that while numeric scores varied considerably, the relative order of submissions remained consistent---suggesting that ranking retains the underlying accuracy of peer evaluations even when absolute scores fluctuate. Caragiannis et al.\@\cite{Caragiannis2020} noted that ranking systems are also less susceptible to strategic manipulation. In settings where students might deliberately assign low scores to peers to elevate their own standing \cite{Jingjing2015}, ranking systems reduce such incentives. Furthermore, ranking helps mitigate the impact of grader variability, such as overly lenient or harsh reviewers.  

Categorization approaches offer several benefits. They come with clear criteria, which help improve reliability. Students follow predefined guidelines and only need to make selections---especially in binary checklists, where they make simple yes/no decisions. Categorization has been shown to be effective in terms of accuracy. Chen et al.\@\cite{Chenwest2022} found that the distribution of student responses in categorical tasks closely matched those of TAs.

Formative feedback is not a direct parameter for grade calculation, but it plays a vital role in ensuring accuracy and fairness. Formative evaluations focus on providing constructive feedback that helps justify the score given to an assignment \cite{DBoud1981}. Gehringer et al.\@\cite{Gehringer2003} noted that inconsistencies in peer reviews are common when numeric grading scales are used alone, without qualitative feedback to provide context. They used formative feedback to ensure fairness by downgrading raters who gave low grades without justification. Formative feedback is often used to ensure transparency \cite{Toll2017}, accountability and fairness \cite{Ou2024}, all of which help build trust among students \cite{kay2022}. Some systems use it to motivate more diligent peer grading, which in turn increases the reliability of the system \cite{Wang2020}.

While summative feedback is noisy and biased, different reviewer weighting strategies have shown promise in mitigating these issues. The consensus-based weighting strategy has proven effective in improving the accuracy of peer grading, particularly in noisy environments, by discounting outliers and unreliable inputs. James et al. \cite{James2017} demonstrated that consensus-based weighting with an Individual Weight Factor (IWF) reduced high-grade errors by over 30\% compared to the naive mean and outperformed both least trimmed squares and Huber estimators. Walsh et al.\@\cite{Walsh2014} showed that consensus-based weighting allowed their system to tolerate up to 25\% bias in peer marking, enhancing fairness. Steinhardt et al.~\cite{Steinhardt2016} further proved that high-quality items could be recovered using only the internal structure of peer consensus, even in the presence of malicious graders. James et al.\@\cite{James2018} reported a higher exact match rate of consensus-based weighted grades to true grades than arithmetic average (60.7\% vs. 56.4\%). 

However, consensus-based weighting is not without limitations. It relies on the assumption that most graders are honest and competent. A key concern is the suppression of rare but correct feedback, as agreement-based weighting can discount an accurate but minority response. Another drawback is its vulnerability to strategic behavior: participants may submit arbitrary and popular grades without effort. To mitigate this, Wang et al.\@\cite{Ros2018} proposed a strategy based on ``surprisingly common'' answers---grades that align with peer consensus on the same assignment but were uncommon across the broader gradebook---were rewarded more.

Calibration-based weighting can help mitigate some of the problems associated with strategic manipulation in consensus-based systems by incorporating instructor-validated ground truth. Systems such as LearnEval and TRUPEQA demonstrate substantial improvements through calibration. LearnEval increased the correlation between peer and instructor grades from 0.19 to 0.72 after calibration\@\cite{Badea2022}, while TRUPEQA consistently achieved lower root mean square error than simple averaging or Gibbs sampling methods with calibration\cite{Prajapati2020}. Calibration also promotes fairness by mitigating collusion and score inflation by penalizing or filtering out reviewers whose grades deviate significantly from known standards\cite{Fang2017}.

Reputation-based weighting has notable strengths in improving grading accuracy across multiple studies. For instance, García-Martínez et al.\@\cite{García-Martínez2019} reported that using engagement and performance-based weighting increased correlation with instructor grades from 0.459 to 0.767. Capuano et al.\@\cite{Capuano2016} showed that reputation-based methods like PowPeerRank significantly reduced RMSE compared to simple averaging. It has also been shown to be effective in bias correction by compensating for consistent over-markers or under-markers\@\cite{Davies2006}. Moreover, reputation-based systems can promote high-quality assessment and accountability by introducing reputation scores or token-based incentives, as demonstrated in the system proposed by Alipour et al.\@\cite{Alipour2024}.

A reputation-based system's effectiveness often depends on the availability of sufficient historical data to accurately estimate grader reliability \cite{Jimenez-Romero2013}. Sajjadi et al.\@\cite{Sajaddi2016} showed that reputation-based grade aggregation beat unweighted mean aggregated scores on synthetic data but not in the real classroom. Moreover, tying grading weight to student performance, such as quiz scores \cite{García-Martínez2019} or prior grades, can risk entrenching privilege and disadvantaging novice learners or those from less-prepared backgrounds, raising equity concerns . The opacity of reputation algorithms may also reduce perceived fairness, especially if students do not understand how their evaluations are being weighted.

Several common issues in consensus and reputation-based systems---such as limited transparency and insufficient motivation for thoughtful reviewing---can be mitigated through back-evaluation based weighting strategy. Staubitz et al.\@\cite{Staubitz2016} demonstrated that 80\% of participants reported feeling more motivated to write useful feedback once they became aware that their reviews would be evaluated. Similarly, Gehringer et al.\@\cite{Gehringer2003} found that students rated the ``review-of-review'' process 3.9 out of 5 in terms of motivating them to write more careful and constructive reviews. This increased transparency not only encouraged accountability but also led to improvements in grading accuracy and perceived fairness. In another study, Toll et al.\@\cite{Toll2017} showed that allowing authors to score the reviews they received motivated reviewers to spend more time and produce higher-quality feedback. Notably, in their study, while 52\% of students initially expected to receive inaccurate scores, only 31\% actually experienced it.

Leveraging additional factors such as formative feedback quality and social relationships among peers can further enhance the accuracy and fairness of peer assessment. For example, Yuan et al.~\cite{Yuan2020} incorporated the thoroughness of textual feedback as a weighting factor in grade aggregation, achieving a mean squared error 6\% lower than other baseline methods. Song et al.\cite{7344292} showed that collusion inflated peer-assessment scores by 1--2 points on average and, in extreme cases, up to 5--10 points. Accounting for such collusive behavior can lead to more accurate and trustworthy grading outcomes.

\subsection{RQ2: Students' Learning Outcomes and Graders' Cognitive Workload}

Assigning ratings can promote critical thinking and engagement among peers \cite{Gracias2013}. However, meaningfully assigning numeric scores to submission quality can be cognitively demanding \cite{Nakayama2022}. Davies et al.\@\cite{Davies2004} reported that students found it difficult to assign accurate numeric scores overall. Kulkarni et al. \@\cite{Kulkarni2013} mentioned that graders felt overwhelmed when required to differentiate between scores on a large-range point scale, which burdens students and increases their cognitive load.  The challenges of cognitive load and grading confusion can be mitigated through strategies such as using smaller-range scales and providing clearer rubric guidelines \cite{Strang2013}. For example, Nakayama et al.\@\cite{Nakayama2022} demonstrated that a 5-point grading scale outperformed a 10-point scale, reducing complexity while maintaining grading effectiveness. 

 Students found ranking less cognitively demanding than scoring. Raman et al.\@\cite{Raman2015} highlighted that ranking reduces the cognitive burden by eliminating the need to interpret an absolute grading scale. Moffat et al.\@\cite{Moffat2020} echoed this, noting that students found it easier to compare peer work relative to one another rather than judge against an abstract universal standard. They also observed that ranking encourages deeper critical engagement and justification, thereby fostering higher-order thinking skills. 

 However, a drawback is when submissions are very similar in quality, students may find it difficult to distinguish between them, leading to frustration and reduced confidence in their judgments \cite{Barber2018}. Moreover, when students are asked to rank a large number of submissions, the task can become overwhelming and may lead to careless or inconsistent rankings \cite{Caragiannis2016, Moffat2020}. One promising solution is pairwise comparison (often called adaptive comparative judgment), where students compare items two at a time. Raman et al.\@\cite{Raman2015} found this approach to significantly ease the grading burden, while Zhang et al.\@\cite{L.Zhang2024} reported that pairwise ranking enabled even novice student reviewers to achieve scoring accuracy comparable to expert lecturers.

Formative feedback helps graders develop their reasoning skills \cite{Albano2017}, foster critical thinking \cite{Caldwell2015}, and even improve their writing skills \cite{Vu2018}. While the grader gains some learning outcomes by grading using summative feedback \cite{Davies2004}, simply assigning a score doesn’t help the author gain much learning and can discourage meaningful reflection \cite{Kulkarni2014}. Gamage et al.\@\cite{Gamage2018} mentioned that grades were viewed as less valuable than feedback, which was often seen as the most valuable part of the assignment. Formative feedback often includes actionable recommendations that help students improve their learning and enhance the quality of submissions~\cite{Andriamiseza2023, deAlfaro2014, Jingjing2015}. However, sometimes students found it more difficult to write detailed comments and justifications than to assign ratings using a scale~\cite{Vu2018}.

 Calibration has been shown to produce significant learning gains when used as a preparatory training step before peer evaluation begins. Caldwell et al.\@\cite{Caldwell2015} found that even weaker students who underwent the training and calibration process were able to accurately and reliably evaluate higher-performing peers, demonstrating cognitive growth and improved understanding of assessment quality. Systems like MTA-2 \cite{Zarkoob2024} and LearnEval \cite{Badea2022} demonstrate that repeated peer assessment under reputation-based frameworks can improve graders’ evaluation skills over time, with measurable increases in dependability and self-reported confidence.

\subsection{RQ3: Achieving Accurate Peer Grades with Minimal Effort}

Most peer grading systems assign multiple graders to each submission. In this context, aggregating rating or category-based responses is typically straightforward and can be accomplished using simple statistical methods such as mean or median \cite{SBennett2012, Kulkarni2013}, or through linear models \cite{HLynda2017, Zarkoob2019}. Numeric scores can also be synthesized using more advanced techniques, including graph-theoretic algorithms \cite{Capuano2016} and Bayesian inference models \cite{Zarkoob2023}. In the ranking-based approach, game theory and graph theory can be applied to derive a final ranking from a large set of elements. For instance, Zhang et al.\@\cite{L.Zhang2024} applied the Swiss tournament system \cite{wikipedia2025swiss} to pair projects with similar rankings in each round, deriving a global ranking from the partial rankings provided by students in an adaptive ranking-based peer grading system. Ling et al.\@\cite{Ling2021} and Caragiannis et al.\@\cite{Caragiannis2014} represent partial rankings as preference graphs, where nodes represent submissions and directed edges represent pairwise preferences. They employ aggregation techniques, including Borda Count \@\cite{wikipedia2025borda} and spectral methods based on Graph Laplacians, to derive a global ranking of peer grades. 

While summative feedback is scalable and can be easily aggregated using various algorithms, evaluating formative feedback remains challenging. Most systems rely on manual approaches---such as back-evaluation or instructor intervention to assess the quality or truthfulness of the feedback. A promising attempt to automate this process was proposed by Yuan et al.~\cite{Yuan2020}, who predicted review thoroughness scores using a large language model (BERT). These scores were then used to assign weights to the corresponding peer grades.

Consensus-based systems are among the most scalable reviewer weighting strategies, requiring neither calibrated assignments from instructors nor historical reputation data. Hamer et al.\@\cite{Hamer2005} emphasized that their consensus-based weighting model handled over 100 essays with no manual calibration, filtering noise algorithmically. Zarkoob et al.\@\cite{Zarkoob2023} demonstrated that their Bayesian models, which use peer consensus as a weighting factor, achieved lower mean absolute errors than those of teaching assistants when more than four peer graders were involved. Like consensus-based systems,  reputation-based systems can also scale effectively, as they rely on automatic weighting with minimal TA intervention~\cite{Zarkoob2024, deAlfaro2014, Walsh2014}. While reputation-based systems are efficient in the long term, their scalability is limited initially by the cost of setup and expert-labeled samples. It requires a non-trivial bootstrapping phase with active learning and manual grading, which can be resource-intensive. Moreover, initial reputation parameters require thoughtful design, especially when few prior grades exist or when switching between domains. 

Calibration is a good choice when no/few prior grades exist, and it is scalable as well. Though it requires instructor intervention or engagement, only a few submissions need to be graded by staff. Fang et al.\@\cite{Fang2017} showed that with just 10 TA-graded submissions, improvements were achieved in a population of 1,000 students. Moreover, another promising approach was proposed by Han et al. \@\cite{Han2023}, who automated the generation of calibrated submissions. Their automated autograder helped reduce grading error by up to 82.77\% \cite{Han2023}. Similarly, Kulkarni et al.\@\cite{Kulkarni2014} used machine-learning text classifiers to predict scores for short-answer responses, offering another way to reduce reliance on manual calibration.

While back-evaluation might add some burden for students, it is highly scalable from the instructor's perspective. Back-evaluation scores can be automatically weighted with a given score by linear equations\@\cite{Sofjan2023}.  When not automated, teachers have to review only disputed or extreme cases ~\cite{Toll2017, Storjohann2015}.

\section{Discussion}
In this section, we reflect on key insights from our findings and highlight areas for future research.

\vspace{3pt}
\noindent\textit{Implications of the Taxonomy:} Our taxonomy provides a structured framework for analyzing and designing peer-grading systems. This classification serves as a descriptive framework and a critical lens to assess the strengths and limitations of various peer-grading approaches across multiple factors such as accuracy, graders learning outcomes, engagement, and scalability. The taxonomy will help educators and system designers identify trade-offs and select appropriate mechanisms for their context.

\vspace{3pt}
\noindent\textit{Toward a Hybrid Model for Effective Peer Grading:} Designing an effective peer-grading system requires a staged, hybrid approach. Calibration and back-evaluation can help train students and establish accountability in the early phases. Once graders are trained, consensus-based weighting can be leveraged to filter noise and stabilize scores. With sufficient historical data, reputation-based weighting can be introduced to enhance accuracy further. Finally, integrating a challenge mechanism can allow students to dispute unfair reviews, reinforcing transparency and closing the feedback loop.

\vspace{3pt}
\noindent\textit{Recommendations and Future Work:} Future peer-grading systems should focus on evaluating formative feedback at scale, using techniques like LLM, which currently receives less attention. While short-term learning outcomes are receiving increasing attention, the long-term effects of peer evaluation on student performance and development need to be explored. Moreover, our taxonomy can be extended to include review-distribution techniques, grade synthesis algorithms such as linear equations, probabilistic models, graph theory, and game-theoretic approaches, along with their implications.

\vspace{3pt}
\noindent\textit{Limitations:} This review may overlook valuable insights from literature and studies not in English or not indexed in the selected databases. The amount of research in peer grading is but a small fraction of the work in peer assessment overall, and many kinds of peer-assessment results might enhance a discussion of peer grading. Further, while comprehensive, our taxonomy may require adaptation for specific contexts. It focuses on evaluation and weighting strategies but does not deeply address other important factors, such as reviewer assignment or incentive structures. 

\section{Conclusion}
This paper presents a comprehensive taxonomy of peer grading systems, organizing them along two key dimensions. By synthesizing findings across diverse studies, we highlight how different design choices impact core outcomes such as accuracy, reliability, fairness, learning, engagement, and scalability. Our analysis shows that no single approach is universally optimal---effective peer grading requires hybrid, staged strategies tailored to the context and maturity of the system. We also identify gaps in current research, including the need for scalable formative feedback evaluation, dynamic weighting, and long-term impact analysis. We hope this work serves as a foundation for designing more robust, transparent, and pedagogically meaningful peer grading systems.

\bibliographystyle{IEEEtran}

\bibliography{references}

\end{document}

%% file: prisma.tex
\tikzset{
    mynode/.style={
        draw, rectangle, align=left, text width=6cm, font=\Large, inner sep=3ex},
    mylabel/.style={
        draw, rectangle, align=center, rounded corners, font=\huge\bf, inner sep=3ex, 
        fill=cyan!30, minimum height=3.8cm},
    arrow/.style={
        very thick,->,>=stealth}
}

\begin{tikzpicture}[
    node distance=3.2cm,
    start chain=1 going below,
    every join/.style=arrow,
    ]

    \coordinate[on chain=1] (tc);

    \node[mynode, on chain=1] (n1)
    {\textbf{742} studies imported for screening\\
    \textbf{352} ACM\\
    \textbf{163} Web of Science\\
    \textbf{79} EBSCO\\
    \textbf{62} ERIC\\
    \textbf{34} PsycINFO\\
    \textbf{32} IEEE Xplore\\
    \textbf{20} PubMed};

    \node[mynode, join, on chain=1] (n2)
    {\textbf{439} studies screened};

    \node[mynode, join, on chain=1] (n3)
    {\textbf{134} full-text studies assessed for eligibility};

    \node[mynode, join, on chain=1] (n4)
    {\textbf{122} studies included};

    \begin{scope}[start chain=going right]
        \chainin (n1);
        \node[mynode, join, on chain, text width=5cm]
        {\textbf{16} duplicates identified manually\\
        \textbf{287} duplicates identified by Covidence \cite{Covidence} };

        \chainin (n2);
        \node[mynode, join, on chain, text width=5cm]
        {\textbf{305} studies irrelevant};

        \chainin (n3);
        \node[mynode, join, on chain, text width=5cm]
        {\textbf{12} studies excluded\\
        };
    \end{scope}

    \begin{scope}[start chain=going below, xshift=-6.5cm, node distance=1cm]
        \node[mylabel, on chain, yshift=-6.2cm, minimum height=8cm] {\rotatebox{90}{Identification}};
        \node[mylabel, on chain, ] {\rotatebox{90}{Screening}};
        \node[mylabel, on chain] {\rotatebox{90}{Eligibility}};
        \node[mylabel, on chain] {\rotatebox{90}{Included}};
    \end{scope}

    
\end{tikzpicture}

%% file: peer_grading_diagram.tex
\definecolor{levelOneColor}{HTML}{FFD6DA}   
\definecolor{levelTwoColor}{HTML}{C9C3F5}   
\definecolor{levelThreeColor}{HTML}{6EE1A0} 
\definecolor{levelFourColor}{HTML}{C3F7D4}  

\begin{forest}
  for tree={
    shape=rounded rectangle,
    rounded corners=10pt,
    thin,
    draw=none,         
    edge={->},
    align=center,
    font=\sffamily,
    inner sep=5pt,
  },
  level 1/.style={
  shape=circle,
  fill=levelOneColor,
  minimum size=10em   
},
  [Peer Grading System, fill= cyan!20
    [Evaluation Approaches, fill=levelTwoColor
      [Summative, fill=levelThreeColor
        [Rating,         fill=levelFourColor]
        [Ranking,        fill=levelFourColor]
        [Categorization, fill=levelFourColor]
      ]
      [Formative, fill=levelThreeColor]
    ]
    [Reviewer Weighting Strategies, fill=levelTwoColor
        [Calibration,     fill=levelThreeColor]
        [Back Evaluation, fill=levelThreeColor]
        [Consensus,       fill=levelThreeColor]
        [Reputation,      fill=levelThreeColor]
    ]
  ]
\end{forest}